\documentclass[11pt,reqno]{amsart}
\usepackage{amscd,amssymb,amsmath,amsthm}
\usepackage[arrow,matrix]{xy}
\usepackage{graphicx}
\usepackage{cite}
\newtheorem{thm}{Theorem}[section]
\newtheorem{defn}[thm]{Definition}
\newtheorem{lemma}[thm]{Lemma}
\newtheorem{pro}[thm]{Proposition}
\newtheorem{rk}[thm]{Remark}

\numberwithin{equation}{section} 

\newcommand{\bea}{\begin{eqnarray}}
\newcommand{\eea}{\end{eqnarray}}

\newcommand{\Z}{\mathbb{Z}}

\newcommand{\R}{\mathbb{R}}
\newcommand{\C}{\mathbb{C}}

\def\R{\mathbb{R}}

\def\C{\mathbb{C}}

\def\Z{\mathbb{Z}}

\begin{document}
\title[Solutions of a Discrete Schr\"odinger Equation]{Periodic Solutions of Discrete Generalized
Schr\"odinger Equations on Cayley Trees}

\author{Fumio Hiroshima,  J\'ozsef L\H{o}rinczi,  Utkir Rozikov}

\address{F. Hiroshima\\ Faculty of Mathematics, Kyushu University, Fukuoka, 819-0395,   Japan}
\email {hiroshima@math.kyushu-u.ac.jp}

 \address{J. L\H{o}rinczi\\ Department of Mathematics, Loughborough University, LE11 3TU, UK}
\email {J.Lorinczi@lboro.ac.uk}

 \address{U.\ A.\ Rozikov\\ Institute of mathematics,
29, Do'rmon Yo'li, 100125, Tashkent, Uzbekistan}
\email {rozikovu@yandex.ru}

\begin{abstract} In this paper we define a discrete generalized Laplacian with arbitrary real power on a
Cayley tree. This Laplacian is used to define a discrete generalized Schr\"odinger operator on the tree.
The case discrete fractional Schr\"odinger operators with index  $0 < \alpha <  2$ is considered in detail,
and periodic solutions of the corresponding fractional Schr\"odinger equations are described. This periodicity
depends on a subgroup of a group representation of the Cayley tree. For any subgroup of finite index we give a
criterion for eigenvalues of the Schr\"odinger operator under which periodic solutions exist. For a normal
subgroup of infinite index we describe a wide class of periodic solutions.
\end{abstract}
\maketitle



\vspace{2cm}

\section{Introduction}

The spectral properties of Schr\"odinger operators on graphs have many applications in physics and they have
been intensively studied since the late 1990s. In \cite{Kl} Klein proved the existence of a purely absolutely
continuous spectrum on the Cayley tree, under weak disorder.

The paper \cite{Mil} reviews the main aspects and problems in the Anderson model on the Cayley tree. It shows
whether wave functions are extended or localized is related to the existence of complex solutions of a certain
non-linear equation. In \cite{K} sufficient conditions for the presence of the absolutely continuous spectrum
of a Schr\"odinger operator on a rooted Cayley tree are given; see \cite{Der} for more results on the model. In
\cite{HS} the spectrum of a Schr\"odinger operator on the $d$-dimensional lattice $\mathbb Z^d$ is studied.

In recent years, fractional calculus has received a great deal of attention. Equations involving fractional
derivatives and fractional Laplacians have been studied by various authors, see e.g. \cite{Po,H,BDST} and the references
therein. In probability theory fractional Laplacians appear as generators of stable processes \cite{KL1}. Non-local (such
as fractional) Schr\"odinger operators are currently increasingly studied \cite{HIL,HLo,KL2,KL3,LM,LKD}. Their counterparts
on lattices are currently much less understood than local discrete Schr\"odinger operators. In \cite{HL} a new phenomenon
on the spectral edge behaviour of non-local lattice Schr\"odinger operators with a $\delta$-interaction has been reported.
In this paper we study solutions of generalized Schr\"odinger equations on Cayley trees.

The paper is organized as follows. In Section 2 we define a discrete generalized Laplacian (with power $\alpha\in \R$)
 on a Cayley tree. This Laplacian is then used to define a discrete generalized Schr\"odinger operator on the Cayley tree.
In Section 3 we consider fractional Schr\"odinger operators with exponents $0\leq \alpha\leq 2$, and describe periodic
solutions of the corresponding fractional Schr\"odinger equations. This periodicity depends on a subgroup of a group
representation $G_k$ of the Cayley tree, i.e., for a given subgroup $\widehat G$ of $G_k$ we define a notion of a
$\widehat G$-periodic solution. Related ideas of $\widehat G$-periodic Gibbs measures and periodic $p-$harmonic functions
have been considered in \cite{Ro0,Ro1,RI} before. The notion of $\widehat G$-periodic solutions can, however, be considered
for arbitrary graphs and groups. For any subgroup of finite index we give a criterion for eigenvalues of the fractional
Schr\"odinger operator under which there are periodic solutions. Also, we describe a wide class of periodic solutions for
a normal subgroup of infinite index.

\bigskip

\section{A discrete fractional Laplacian on Cayley trees}

Recall that a \emph{Cayley tree} $\Gamma^k$ is an infinite $k$-regular tree ($k\geq 1$), i.e., a
connected graph on a countably infinite set of vertices, with no cycles and in which every vertex
has degree $k+1$. We denote $\Gamma^k=(V, L)$, where $V$ is the vertex set and  $L$ the edge set
of $\Gamma^k$. Two vertices $x$ and $y$ are called \emph{adjacent} if there exists an edge $l \in L$
connecting them. In such a case we will use the notation $l=\langle x,y\rangle$. Also, we denote the
set of adjacent vertices of $x$ by $S(x)=\{y\in V:  \langle x,y\rangle \in L\}$ and call it the
\emph{neighbourhood} of $x$.

An alternating sequence $(x_1, \langle x_1,x_2 \rangle, x_2, ..., x_{n-1}, \langle x_{n-1}, x_n\rangle, x_n)$
of vertices and edges is a \emph{path}. The \emph{length} of a path is the number of edges connecting the two
end-vertices of the path. On $\Gamma^k$ one can define a natural \emph{distance} $d(x,y)$ as the length of the
shortest path connecting $x$ and $y$.

For any $x\in V$ denote
$$
W_m(x)=\{y\in V: d(x,y)=m\}, \ \ m\geq 1.
$$
Let $u:V\to \C$ be a function. The\emph{ Laplacian} $\Delta$ on a Cayley tree is defined by
\begin{equation}\label{d21}
(\Delta u)(x) =\sum_{y\in S(x)}(u(y)-u(x)).
\end{equation}

For $n \in \Bbb N$ we denote by $\Delta^n$ the $n$-fold iteration of $\Delta$, i.e., $\Delta^n=\Delta(\Delta^{n-1})$.
We have the following expression.

\begin{lemma}
\label{ln}
For any $n \in \Bbb N$ and $x\in V$ we have
\begin{eqnarray}
\label{dn}
\lefteqn{
(\Delta^nu)(x) =} \\
&&
\sum_{j=0}^{n-1}\left[(-(k+1))^j{n\choose j}\sum_{y_1\in S(x)}\sum_{y_2\in S(y_1)}\dots
\sum_{y_{n-j}\in S(y_{n-j-1})}u(y_{n-j})\right]
+(-1)^n(k+1)^nu(x) \nonumber,
\end{eqnarray}
with $y_0=x$.
\end{lemma}
\begin{proof}
We proceed by induction. For $n=1$ this is true by (\ref{d21}). For $n=2$ using (\ref{d21}) we obtain
\begin{eqnarray*}
(\Delta^2u)(x)=(\Delta\Delta u)(x)
&=&
\sum_{y\in S(x)}((\Delta u)(y)-(\Delta u)(x)) \\
&=&
\sum_{y_1\in S(x)}\sum_{y_2\in S(y_1)}u(y_2)-2(k+1)\sum_{y_1\in S(x)}u(y_1)+(k+1)^2u(x).
\end{eqnarray*}
Now we make the induction hypothesis that (\ref{dn}) is true for $n$, and prove it for $n+1$:
\begin{eqnarray}
\label{mi}
\lefteqn{
(\Delta^{n+1}u)(x)=(\Delta\Delta^{n} u)(x) = \sum_{y\in S(x)}((\Delta^n u)(y)-(\Delta^n u)(x))} \nonumber \\
&=&
\sum_{y\in S(x)}\left(\sum_{j=0}^{n-1}(-(k+1))^j{n\choose j} 
\sum_{y_1\in S(x), \, y_2\in S(y_1), \ldots \atop y_{n-j}\in S(y_{n-j-1})}u(y_{n-j}) +(-(k+1))^nu(y) \right.  \nonumber \\
&&
\qquad \left.-\sum_{j=0}^{n-1}\left[(-(k+1))^j{n\choose j}
\sum_{y_1\in S(x), \, y_2\in S(y_1), \ldots \atop y_{n-j}\in S(y_{n-j-1})}u(y_{n-j})\right]+
 (-(k+1))^nu(x) \right) \nonumber \\
&=&
\sum_{y\in S(x)}\sum_{y_1\in S(y)}\sum_{y_2\in S(y_1)}\dots\sum_{y_{n}\in S(y_{n-1})}u(y_{n}) \nonumber \\
&&
\quad +\sum_{j=1}^{n-1}\left[(-(k+1))^j{n\choose j}\sum_{y\in S(x), y_1\in S(y) \atop y_2\in S(y_1) \dots
y_{n-j}\in S(y_{n-j-1})}u(y_{n-j})\right]  \nonumber \\
&&
\quad -(k+1)\sum_{j=0}^{n-1}\left[(-(k+1))^j{n\choose j}\sum_{y_1\in S(x), y_2\in S(y_1) \dots \atop y_{n-j}\in S(y_{n-j-1})}
u(y_{n-j})\right] \nonumber \\
&&
\quad  +(-(k+1))^n\sum_{y\in S(x)}u(y) -(k+1)(-(k+1))^n{n\choose n}u(x).
\end{eqnarray}
Relabelling the variables $y$ by $y_1$ and $y_i$ by $y_{i+1}$ for $i=1,\dots,n$, and using   ${n+1\choose j}=
{n \choose j-1}+{n\choose j}$, we obtain
\begin{eqnarray*}
\lefteqn{
{\rm RHS \ \ of}\ \ (\ref{mi}) =
\sum_{y_1\in S(x)}\sum_{y_2\in S(y_1)}\dots\sum_{y_{n+1}\in S(y_{n})}u(y_{n+1}) +(-(k+1))^{n+1}u(x) } \\
&&
+\sum_{j=1}^n\left[(-(k+1))^j{n\choose j}\sum_{y_1\in S(x), y_2\in S(y_1) \dots \atop y_{n-j+1 \in S(y_{n-j})}}
u(y_{n-j+1})\right] \\
&&
+\sum_{j'=1}^{n}\left[(-(k+1))^{j'}{n\choose j'-1}\sum_{y_1\in S(x), y_2\in S(y_1) \dots \atop
y_{n-j'+1}\in S(y_{n-j'})}u(y_{n-j'+1})\right]  \\
&=&
\sum_{y_1\in S(x), y_2\in S(y_1) \dots \atop y_{n+1}\in S(y_{n})}u(y_{n+1}) +(-(k+1))^{n+1}u(x)  \\
&&
+ \sum_{j=1}^n\left[(-(k+1))^j\left({n\choose j}+{n\choose j-1}\right)\sum_{y_1\in S(x), y_2\in S(y_1)\dots
\atop y_{n-j+1}\in S(y_{n-j})}u(y_{n-j+1})\right] \\
&=&
\sum_{j=0}^{n}\left[(-(k+1))^j{n+1\choose j}\sum_{y_1\in S(x), y_2\in S(y_1) \dots
\atop y_{n-j+1}\in S(y_{n-j})}u(y_{n-j+1})\right]+(-(k+1))^{n+1}u(x).
\end{eqnarray*}
\end{proof}

It is easily seen that
\begin{equation}
\label{ss}
\sum_{y_1\in S(x)}\sum_{y_2\in S(y_1)}u(y_{2}) = \sum_{y\in W_2(x)}u(y)+(k+1)u(x).
\end{equation}

Next we define real powers $\Delta^\alpha$ of the Laplacian in such a way that when $\alpha$ takes a positive integer
value $n$, then $\Delta^n$ is recovered. To define a fractional Laplacian on the Cayley tree we make the change
$$
{\alpha \choose j} \mapsto {\Gamma(\alpha+1)\over \Gamma(j+1)\Gamma(\alpha+1-j)}.
$$
Note that this is the multi-dimensional version of the discrete fractional difference operator introduced
by Andersen for a sequence $(u_n)_{n\in \Bbb N}$ of real numbers (see \cite{H})
$$
\Delta^\alpha u_n = \sum^\infty_{j=0}(-1)^j{\alpha\choose j}u_{n+j}.
$$
\begin{defn}[\textbf{Generalized Laplacian on a Cayley tree}]
We define the\emph{ generalized Laplacian} of order $\alpha\geq 0$ on a Cayley tree by
\begin{eqnarray}
\label{da}
(\Delta^\alpha u)(x)
&=&
\sum_{j=0}^{[\alpha]-1}\left[A(\alpha,j)\sum_{y_1\in S(x)}\sum_{y_2\in S(y_1)}\dots
\sum_{y_{[\alpha]-j}\in S(y_{[\alpha]-j-1})}u(y_{[\alpha]-j})\right]  \nonumber\\
&& \qquad + (-(k+1))^{[\alpha]}u(x),
\end{eqnarray}
where $A(\alpha, j)={(-(k+1))^j\Gamma(\alpha+1)\over \Gamma(j+1)\Gamma(\alpha+1-j)}$ and $[\alpha]$ denotes
the integer part of $\alpha$. When $0 < \alpha < 2$, we call $\Delta^\alpha$ \emph{fractional Laplacian} of
order $\alpha$.
\end{defn}

A Schr\"odinger equation of a fractional power $\alpha$ can be written as
\begin{equation}\label{saa}
(\Delta^\alpha u)(x)=Eu(x)-v(x)u(x).
\end{equation}
Here $u(x)\in \C$ is the value of the wave function at vertex $x$, $E\in \R$ is the energy of the
particle, and $v=\{v(x)\}_{x\in V}$ is a given real-valued function.

\bigskip

\section{Periodic wave functions}

In what follows we consider the case $0\leq\alpha \leq 2$.

\subsection{\bf Case: $0\leq\alpha<1$}  We have by (\ref{da})
\begin{equation}\label{du1}
(\Delta^\alpha u)(x)=(-(k+1))^{[\alpha]}u(x)=u(x),
\end{equation}
i.e., the operator $\Delta^\alpha$ is an identity operator.
Consequently, from (\ref{saa}) we obtain
 \begin{equation}\label{saa1}
u(x)=Eu(x)-v(x)u(x).
\end{equation}
From this equation we get $(E-1-v(x))u(x)=0$. A straightforward analysis of this equation gives the following.

\begin{pro}
For  given $E$ and function $v$ the equation (\ref{saa1}) has the solution
$$
u(x)=\left\{\begin{array}{ll}
0, \ \ \ \ \ \ \ \mbox{if} \ \ x\in\{y: v(y)+1\ne E\}\\[2mm]
u_*(x), \ \ \mbox{if} \ \ x\in\{y:  v(y)+1=E\},
\end{array}
\right.$$
where $u_*$ is an arbitrary function.
\end{pro}

\begin{rk}
{\rm
If one wants to have a fractional Laplacian $\Delta_1^\alpha$ with the property that it is not identical for any
$\alpha>0$, then this operator can be defined using operator (\ref{da}) as
\begin{equation}\label{01}
(\Delta_1^\alpha u)(x)=\left\{\begin{array}{ll}
(\Delta^{\alpha+1} u)(x), \ \ \mbox{if} \ \ \alpha\in [0,+\infty)\setminus \{0,1,2,\dots\}\\[3mm]
(\Delta^{\alpha} u)(x), \ \ \mbox{if} \ \ \alpha\in \{0,1,2,\dots\}.
\end{array}
\right.
\end{equation}
The Schr\"odinger equation (\ref{saa}) corresponding to the operator $\Delta_1^\alpha$ is
\begin{equation}\label{02}
(\Delta_1^\alpha u)(x)=Eu(x)- v(x)u(x).
\end{equation}
We note that if we know the solutions of the equation (\ref{saa}), then using (\ref{01}) we can find solutions of
(\ref{02}). Thus it is sufficient to consider the equation (\ref{saa}) for the operator $\Delta^\alpha$.
}
\end{rk}

\subsection{\bf Case: $\alpha=1$} In this case
the Schr\"odinger equation reads
\begin{equation}\label{hr}
\sum_{y\in S(x)}u(y)=Eu(x)- v(x)u(x).
\end{equation}

\subsubsection{Properties of group representations of the Cayley tree}
\begin{figure}
  \includegraphics[width=11cm]{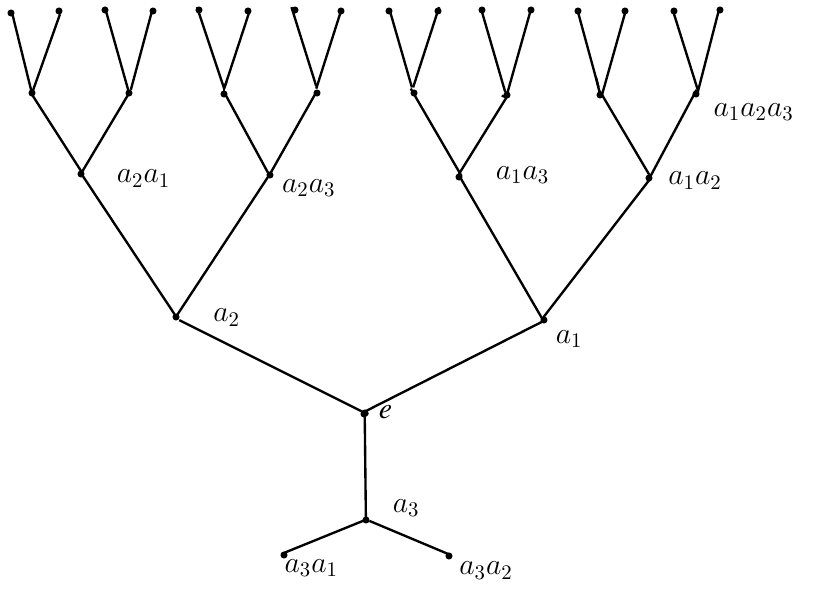}\\
  \caption{Cayley tree of order two, i.e., $k=2$.}\label{fig1}
\end{figure}

To study solutions of equation (\ref{hr}) and equations considered below first we give some properties of a group
representation of the Cayley tree. Let $G_k$ be a free product of $k + 1$ cyclic groups of the second order with
generators $a_1, a_2,\dots, a_{k+1}$, respectively. Any element $x\in G_k$ has the following form:
$$
x=a_{i_1}a_{i_2}\dots a_{i_n},\ \ \mbox{where} \ \ 1\leq i_m\leq k+1,\, m= 1,\dots, n.
$$
It is known \cite{Ga0} that there exists a one-to-one correspondence between the set of vertices $V$ of the Cayley
tree $\Gamma^k$ and the group $G_k$. In Fig. \ref{fig1} some elements of $G_k$ are shown (see \cite[Chapter 1]{Ro} for detailed
properties of this group representation).

Let $K\subset G_k$ be an arbitrary normal subgroup of index $r$ of the group $G_k$ (see \cite{Ga3,Ro,Ro1} for examples
of subgroups of $G_k$). We introduce the following equivalence relation on the set $G_k$: $x\sim y$ if $xy^{-1}\in K$.
Let $G_k/K = \{K_1, K_2,\dots, K_{r}\}$ be the factor-group with respect to $K$. Denote
$$
q_i(x) = |W_1(x)\cap K_i|, \ \ s_i(x)=|W_2(x)\cap K_i|, \ \ i=1,\dots, r, \ \ x\in G_k,
$$
where $|\cdot|$ is the counting measure of a set. The following lemma is known \cite{N,Ro0}.

\begin{lemma}\label{ue}
If $x\sim y$, then $q_i(x)=q_i(y)$ and $s_i(x)=s_i(y)$ for any $i = 1,\dots, r$.
\end{lemma}

By Lemma \ref{ue} it follows that $q_j(x)$ and $s_j(x)$ corresponding to $K$ have the following form
$$
q_j(x)=q_{ij}, \ \ s_j(x)=s_{ij}\ \ \mbox{for all} \ \  x\in K_i.
$$
Since the set $V$ of vertices has the group representation $G_k$, without loss of generality we identify $V$ with $G_k$,
i.e., we may replace $V$ with $G_k$.

\begin{defn}\label{d1}
Let $K$ be a subgroup of $G_k, k\geq 1$. A function $u:x\in G_k\to u(x)\in\C$ is called {\rm $K$-periodic} if $u(yx)=u(x)$
for all $x\in G_k$ and $y\in K$. A $G_k$-periodic function $u$ is called {\rm translation-invariant}.
\end{defn}

\subsubsection{The case of finite index}\label{sf}

Let $G_k/K=\{K_1,...,K_r\}$  be a factor group, where $K$ is a normal subgroup of index $r\geq 1$. Note that any $K$-periodic
function $u$ has the form
\begin{equation}\label{pfa}
u(x)=u_j, \ \ \forall x\in K_j,\ \ j=1,\dots,r.
\end{equation}
Assume that $v=\{v(x)\}_{x\in V}$ is $K$-periodic, i.e. $v(x)=v_j$ for any $x\in K_j$ then from (\ref{hr}) we get
\begin{equation}\label{psa}
\sum_{j=1}^rq_{ij}u_j=(E- v_i)u_i, \ \ i=1,2,\dots,r.
\end{equation}
This system can be written as
\begin{equation}\label{ps1a}
 \sum_{{j=1\atop j\ne i}}^rq_{ij}u_j-(E-q_{ii}- v_i)u_i=0, \quad i=1, \ldots, r.
\end{equation}
For a given normal subgroup $K$ of index $r\geq 1$ we denote by $D_K(E, v)$ the determinant of the system (\ref{ps1a}).
Thus the following holds.

\begin{thm}
For given $K$ and a $K$-periodic $v$ there exists a $K$-periodic wave function $u\ne 0$ if and only if $E$ is a solution to
$D_K(E, v)=0$.
\end{thm}

\noindent
{\bf Examples.}

\noindent
(1) In case $K=G_k$, we have translation invariant wave functions, i.e., $u(x)=u_1$ for all $x\in V$. In this case
$D_K(E, v)=E- v_1-k-1=0$, thus a translation invariant wave function exists if and only if $E= v_1+k+1.$

\medskip
\noindent
(2) Let $K=G^{(2)}_k$ be the subgroup in $G_k$ consisting of all words of even length. Clearly, it is a subgroup of index 2.
In this case we have $D_K(E, v)=(E- v_1)(E- v_2)-(k+1)^2=0$. For any $E$ satisfying this equation there are $G^{(2)}_k$-periodic
wave functions having the form
$$
u(x)=\left\{\begin{array}{ll}
u_1, \ \ \mbox{if} \ \ x\in G^{(2)}_k\\[2mm]
u_2, \ \ \mbox{if} \ \ x\in G_k\setminus G^{(2)}_k.
\end{array}\right.
$$

\begin{figure}
  \includegraphics[width=11cm]{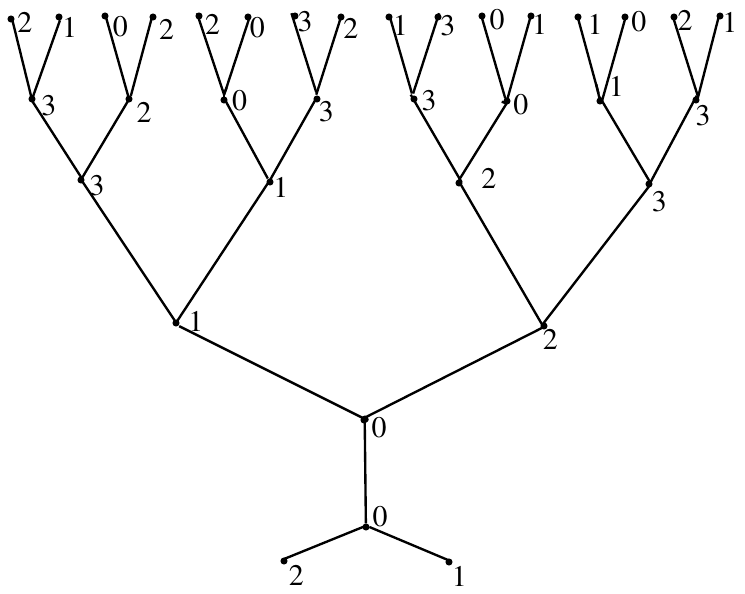}\\
  \caption{Partition of $\Gamma^2$ by subgroup $H_0$ of index 4.}\label{fig2}
\end{figure}

\medskip
\noindent
(3) Let $k=2$, $K=H_0$, with $H_0=\{x\in G_2: \omega_x(a_1)-{\rm even}, \, \omega_x(a_2)-{\rm even}\}$, where the number of letters
$a_i$ of the word $x$ is denoted by $\omega_x(a_i)$. Note that $H_0$ is a normal subgroup of index 4. The group  $G_2/H_0 =
\{H_0, H_1,H_2, H_3\}$ has the following elements:
\begin{eqnarray*}
&& H_1=\left\{x\in G_2: \omega_x(a_1)-{\rm even}, \, \omega_x(a_2)-{\rm odd}\right\} \\
&& H_2=\left\{x\in G_2: \omega_x(a_1)-{\rm odd}, \, \omega_x(a_2)-{\rm even}\right\} \\
&& H_3=\left\{x\in G_2: \omega_x(a_1)-{\rm odd}, \, \omega_x(a_2)-{\rm odd}\right\}.
\end{eqnarray*}
In Fig. \ref{fig2} the partitions of $\Gamma^2$ with respect to $H_0$ are shown. The elements of the class $H_i$, $i=0,1,2,3$ are
denoted by $i$. In this case an $H_0$-periodic wave function $u$ has the form $u(x)=u_i$ if $x\in H_i$. Such a non-zero function
exists if and only if $E$ satisfies the equation
$$
D_{H_0}(E, v)=\left|\begin{array}{cccc}
 v_0+1-E & 1 & 1 &0 \\[2mm]
1&  v_1+1-E & 0 & 1 \\[2mm]
1& 0&  v_2+1-E & 1  \\[2mm]
0& 1 & 1 &  v_3+1-E
\end{array}\right|=0.
$$

\subsubsection{The case of infinite index}\label{si}
\begin{figure}
   \includegraphics[width=12cm]{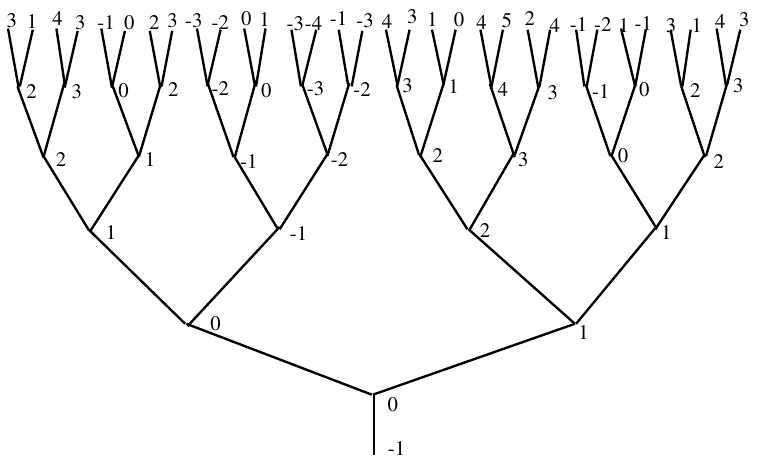}\\
  \caption{The partition of the Cayley tree $\Gamma^2$ w.r.t. $\mathcal H_0$, the elements of the class $\mathcal H_i$,
  $i\in \Z$, are denoted by $i$.}\label{fig3}
\end{figure}

We consider a normal subgroup $\mathcal H_0$ of infinite index constructed for $M=\{1,2\}$ as follows. Let the mapping
$\pi_M:\{a_1,...,a_{k+1}\}\longrightarrow \{a_1,a_2\}\cup \{e\}$ be defined by
$$
\pi_M(a_i)=\left\{%
\begin{array}{ll}
    a_i, & \hbox{if} \ \ i\in M \\
    e, & \hbox{if} \ \ i\notin M. \\
\end{array}
\right.
$$
Denote by $G_M$ the free product of cyclic groups $\{e,a_i\}, \ i\in M$. Consider
$$
f_M(x)=f_M(a_{i_1}a_{i_2}...a_{i_m})=\pi_M(a_{i_1})\pi_M(a_{i_2})...\pi_M(a_{i_m}).
$$
Then it is seen that $f_M$ is a homomorphism and hence $\mathcal H_0=\{x\in G_k: \ f_M(x)=e\}$ is a normal subgroup of
infinite index. The factor group has the form
$$
G_k/\mathcal H_0=\{\mathcal H_0, \mathcal H_0(a_1), \mathcal H_0(a_2), \mathcal H_0(a_1a_2), \dots\},
$$
where $\mathcal H_0(y)=\{x\in G_k: f_M(x)=y\}$. With the notations
$$
\mathcal H_m=\mathcal H_0(\underbrace{a_1a_2\dots}_m),\ \  \ \mathcal H_{-m}=\mathcal H_0(\underbrace{a_2a_1\dots}_m)
$$
the factor group can be represented as
$$
G_k/\mathcal H_0=\{\dots, \mathcal H_{-2}, \mathcal H_{-1}, \mathcal H_0, \mathcal H_1, \mathcal H_2, \dots\}.
$$
The partition of the Cayley tree $\Gamma^2$ with respect to $\mathcal H_0$ is shown in Fig. \ref{fig3} (the elements of
the class $\mathcal H_i$, $i\in \Z$, are denoted by $i$).

Any $\mathcal H_0$-periodic wave function has the form
$$
u(x) = u_m \ \ \mbox{if} \ \ x\in \mathcal H_m, \ \ m\in \Z.
$$
We note that (see \cite{RI}) if $x\in \mathcal H_m$, then
$$
q_{m-1}(x)=|W_1(x)\cap \mathcal H_{m-1}|=1,  \ \ q_m(x)=k-1, \ \ q_{m+1}(x)=1.
$$
Using these equalities we get (see Fig. \ref{fig3})
\begin{eqnarray*}
&& s_{m-2}(x)=|W_2(x)\cap \mathcal H_{m-2}|=1, \ \ s_{m-1}(x)=2(k-1) \\
&& s_{m}(x)=(k-1)(k-2), \ \ s_{m+1}(x)=2(k-1), \ \ s_{m+2}(x)=1.
\end{eqnarray*}
Assuming that $v$ is $\mathcal H_0$-periodic, i.e., $v(x)=v_m$ for any $x\in \mathcal H_m$, equation (\ref{hr}) gives (see Fig. \ref{fig3})
\begin{equation}\label{s3a}
(E- v_n)u_n=u_{n-1}+(k-1)u_n+u_{n+1},
\end{equation}
where $n\in \Z$.
For simplicity, assume that $v_n\equiv v_0$, i.e., $v$ is translation invariant. Then equation (\ref{s3a}) becomes a linear homogeneous
recurrence equation of order 2, i.e.
\begin{equation}\label{aaa}
u_{n+1}+(k+v_0-E-1)u_n+u_{n-1}=0.
\end{equation}

Recall that the solution procedure of a general linear homogeneous recurrence equation with constant coefficients of order $d$ is
given by
\begin{equation}\label{og}
u_{n}+ c_1u_{n-1}+ c_2u_{n-2}+\cdots+c_du_{n-d} = 0.
\end{equation}
Let $p(t)$ be the characteristic polynomial, i.e.,
$$
p(t)=t^d + c_1t^{d-1} + c_2t^{d-2}+\cdots+c_{d}.
$$
Suppose $\lambda_i$ is a root of $p(t)$ having multiplicity $r_i$. Write $u_n$ as a linear combination of {\it all} the roots counting
multiplicity with arbitrary coefficients $b_{ij}$:
\begin{equation}\label{gs}
u_n =\sum_{i} \left(b_{i1}+ b_{i2}n+ b_{i3}n^2+\cdots+b_{ir_i} n^{r_i-1}\right)\lambda_i^n.
\end{equation}
This gives the general solution to equation (\ref{og}). Applying this procedure to equation (\ref{aaa}), we have
\begin{equation}\label{lle}
p(\lambda)=\lambda^2+(k-1+ v_0-E)\lambda+1=0.
\end{equation}
An analysis of this equation by using formula (\ref{gs}) yields the following result.
\begin{thm}
For $\mathcal H_0$ and a translation invariant $v$ the following hold:
\begin{itemize}
\item[(1)]
If $E\in (k+ v_0-3, \, k+ v_0+1)$, then an $\mathcal H_0$-periodic wave function is $u(x)=C_1\lambda_1^n+C_2\lambda_2^n$
for any $x\in \mathcal H_n$, $n\in \Z$, where $\lambda_1, \lambda_2$ are (complex) solutions to equation (\ref{lle}).

\item[(2)]
If $E=k+ v_0-3$, then there is an $\mathcal H_0$-periodic function $u(x)=(-1)^n(C_1+nC_2)$, $x\in \mathcal H_n$, $n\in \Z$.

\item[(3)]
If $E=k+ v_0+1$, then there is an $\mathcal H_0$-periodic function $u(x)=C_1+nC_2$, $x\in \mathcal H_n$.

\item[(4)]
If $E\in (-\infty, k+ v_0-3)\cup(k+ v_0+1, +\infty)$, then an $\mathcal H_0$-periodic wave function is  $u(x)=C_1\lambda_1^n
+C_2\lambda_2^n$ for any $x\in \mathcal H_n$, $n\in \Z$, where $\lambda_1, \lambda_2$ are (real) solutions to equation (\ref{lle}).
\end{itemize}
These solutions are in each case unique up to the choice of the constant prefactors.
\end{thm}

\subsection{\bf Case: $1<\alpha<2$} In this case we have from (\ref{da})
$$
(\Delta^\alpha u)(x)=A(\alpha, 0)\sum_{y\in S(x)}u(y)+A(\alpha,1)u(x)=
\sum_{y\in S(x)}u(y)-{(k+1)\Gamma(\alpha+1)\over \Gamma(\alpha)}u(x).
$$
Consequently, from (\ref{saa}) we get an equation similar to equation (\ref{hr}), with $E$ replaced by $E'=E+{(k+1)\Gamma(\alpha+1)
\over \Gamma(\alpha)}$. Thus the analysis of this equation is very similar to the analysis of equation (\ref{hr}).

\subsection{\bf Case: $\alpha=2$}

Using formula (\ref{ss}) for $\alpha=n=2$, we get
\begin{equation}\label{ss1}
(\Delta^2u)(x)=\sum_{z\in W_2(x)}u(z)-2(k+1)\sum_{y\in W_1(x)}u(y)+(k+1)(k+2)u(x).
\end{equation}
The Schr\"odinger equation corresponding to the bi-Laplacian is
\begin{equation}\label{s}
(\Delta^2 u)(x)=Eu(x)- v(x)u(x).
\end{equation}
Now we discuss the periodic solutions of (\ref{s}).

\subsubsection{The case of finite index}

Let $G_k/K=\{K_1,...,K_r\}$  be a factor group, where $K$ is a normal subgroup of index $r\geq 1$. Assume that $v=\{v(x)\}_{x\in G_k}$
is $K$-periodic, i.e., $v(x)=v_j$ for any $x\in K_j$. Then by (\ref{s}) and (\ref{ss1}) we get
\begin{equation}\label{ps}
 \sum_{j=1}^r s_{ij}u_j-2(k+1)\sum_{j=1}^rq_{ij}u_j+(k+1)(k+2)u_i=(E- v_i)u_i, \ \ i=1, \ldots, r.
\end{equation}
This system can be written as
\begin{equation}\label{ps1}
 \sum_{{j=1\atop j\ne i}}^r(s_{ij}-2(k+1)q_{ij})u_j-(E-(k+1)(k+2)+2(k+1)q_{ii}-s_{ii}- v_i)u_i=0, \quad i=1, \ldots, r.
\end{equation}
For a given normal subgroup $K$ of index $r\geq 1$ we denote by ${\mathcal D}_K(E, v)$ the determinant of the system (\ref{ps1}).
Thus we have the following result.

\begin{thm}
For a given $K$ and a $K$-periodic $v$ there exists a $K$-periodic solution $u\ne 0$ to the equation (\ref{s}) if and only if $E$ is
a solution of ${\mathcal D}_K(E, v)=0$.
\end{thm}

\noindent
{\bf Examples.} We consider some examples of subgroups introduced in Subsection \ref{sf}.

\medskip
\noindent
(1) In case $K=G_k$ we have translation invariant solutions, i.e., $u(x)=u_1$ for all $x\in V$. In this case ${\mathcal D}_K(E, v)=
E- v_1=0$. Thus a non-zero, translation invariant wave function exists if and only if $E= v_1.$

\medskip
\noindent
(2) For  $K=G^{(2)}_k$ we have
$$
{\mathcal D}_K(E, v)=(E- v_1)(E- v_2)-2(k+1)^2(2E-(v_1+v_2))=0.
$$
For any $E$ satisfying this equation there exist $G^{(2)}_k$-periodic solutions of equation (\ref{s}).

\medskip
\noindent
(3) Let $K=H_0$. By Fig. \ref{fig2} it is clear that
\begin{eqnarray*}
&& q_{00}=q_{01}=q_{02}=q_{10}=q_{11}=q_{13}=q_{20}=q_{22}=q_{23}=q_{31}=q_{32}=q_{33}=1 \\
&& q_{03}=q_{12}=q_{21}=q_{30}=0 \\
&& s_{ij}=
\left\{\begin{array}{ll}
0, \ \ \mbox{if} \ \ i=j\\[2mm]
2, \ \ \mbox{if} \ \ i\ne j.
\end{array}\right.
\end{eqnarray*}
Hence an $H_0$-periodic function exists if and only if $E$ satisfies the equation
$$
{\mathcal D}_{H_0}(E, v)=
\left|\begin{array}{cccc}
 v_0+6-E & -4 & -4 &2 \\[2mm]
-4&  v_1+6-E & 2 & -4 \\[2mm]
-4& 2&  v_2+6-E & -4  \\[2mm]
2& -4 & -4 &  v_3+6-E
\end{array}\right|=0.
$$

\subsubsection{The case of infinite index}
Consider $\mathcal H_0$ constructed in Subsection \ref{si}. Then from equation (\ref{s}) using (\ref{ss1}) and the above
expressions of $q_m(x)$ and $s_m(x)$ we obtain
\begin{equation}\label{s3}
(E- v_m)u_m=u_{m+2}-4u_{m+1}+6u_{m}-4u_{m-1}+u_{m-2},
\end{equation}
where $m\in \Z$. For simplicity, we assume that $v_n\equiv v_0$, i.e., $v$ is translation invariant. Then equation (\ref{s3})
yields the characteristic equation
\begin{equation}\label{la}
\lambda^4-4\lambda^3+(6+ v_0-E)\lambda^2-4\lambda+1=0.
\end{equation}
We make use again of the general argument (\ref{og})-(\ref{gs}). Denoting $\xi=\lambda+1/\lambda$, from (\ref{la}) we get
\begin{equation}\label{*}
(\xi-2)^2=E- v_0.
\end{equation}
By a simple analysis of (\ref{*}) and then of $\lambda+1/\lambda=\xi$, we obtain the following results.
\begin{itemize}
\item[(1)]
If $E< v_0$, then equation (\ref{la}) has the four distinct complex solutions
$$
\hat\lambda_{1,2}={1\over 2}\left(z\pm \sqrt{z^2-4}\right), \ \ \hat\lambda_{3,4}={1\over 2}\left(\bar z\pm \sqrt{\bar z^2-4}\right),
$$
where $z=2+i\sqrt{v_0-E}$, $\bar z=2-i\sqrt{v_0-E}$.

\vspace{0.1cm}
\item[(2)]
If $E= v_0$, then equation (\ref{la}) has the unique solution $\lambda=1$ with multiplicity 4.

\vspace{0.1cm}
\item[(3)]
If $ v_0<E<16+ v_0$, then equation (\ref{la}) has the two real solutions
\begin{equation}\label{A}
\lambda_{1,2}=\frac{2+\sqrt{E- v_0}\pm \sqrt{(2+\sqrt{E- v_0})^2-4}}{2}
\end{equation}
and the two complex non-real solutions
\begin{equation}\label{tl}
\tilde\lambda_{3,4}=\frac{2-\sqrt{E- v_0}\pm i\sqrt{4-(2-\sqrt{E- v_0})^2}}{2}.
\end{equation}

\vspace{0.1cm}
\item[(4)]
If $E=16+ v_0$, then equation (\ref{la}) has three real solutions
\begin{equation}\label{A3}
\lambda_{1,2}=3\pm 2\sqrt{2}, \ \ \lambda_3=-1,
\end{equation}
where $-1$ has multiplicity 2.

\vspace{0.1cm}
\item[(5)]
If $E>16+ v_0$, then equation (\ref{la}) has the four real solutions $\lambda_{1}$, $\lambda_2$ given by (\ref{A}) and
\begin{equation}\label{B}
\lambda_{3,4}=\frac{2-\sqrt{E- v_0}\pm \sqrt{(2-\sqrt{E- v_0})^2-4}}{2}.
\end{equation}
\end{itemize}
Thus we have proved the following result.
\begin{thm}
For $\mathcal H_0$ and a translation invariant $v$ (i.e., $v\equiv v_0$) the following hold:
\begin{itemize}
\item[(1)]
If $E< v_0$, then there exists an $\mathcal H_0$-periodic function of the form
$$
u(x)=u_m=C_1\hat\lambda_1^m+C_2\hat\lambda_2^m+C_3\hat\lambda_3^m+C_4\hat\lambda_4^m, \quad x\in \mathcal H_m,
\ \ m\in \Z,
$$
where $\hat\lambda_i$, $i=1,2,3,4$ are the complex numbers above.
\item[(2)]
If $E=v_0$, then equation (\ref{s}) has an $\mathcal H_0$-periodic solution
$$
u(x)=u_m=C_0+C_1m+C_2m^2+C_3m^3, \quad x\in \mathcal H_m, \ \ m\in \Z.
$$
\item[(3)]
If $ v_0<E<16+ v_0$, then equation (\ref{s}) has an $\mathcal H_0$-periodic solution
$$
u(x)=u_m=C_1\lambda_1^m+C_2\lambda_2^m+C_3\tilde\lambda_3^m+C_4\tilde\lambda_4, \quad x\in \mathcal H_m,
\ \ m\in \Z,
$$
where $\lambda_1$, $\lambda_2$ are given by (\ref{A}), and $\tilde\lambda_3, \tilde\lambda_4$ are defined in
(\ref{tl}) above.
\item[(4)]
If $E=16+ v_0$, then equation (\ref{s}) has an $\mathcal H_0$-periodic solution
$$
u(x)=u_m=C_1(3-2\sqrt{2})^m+C_2(3+2\sqrt{2})^m+(C_3+mC_4)(-1)^m,
$$
for all $x\in \mathcal H_m$ and  $m\in \Z$.
\item[(5)]
If $E>16+ v_0$, then equation (\ref{s}) has an $\mathcal H_0$-periodic solution
$$
u(x)=u_m=C_1\lambda_1^m+C_2\lambda_2^m+C_3\lambda_3^m+C_4\lambda_4^m, \quad x\in \mathcal H_m, \ \ m\in \Z,
$$
where $\lambda_1$, $\lambda_2$, $\lambda_3$, $\lambda_4$ are given by (\ref{A}) and (\ref{B}).
\end{itemize}
These solutions are in each case unique up to the choice of the constant prefactors.
\end{thm}

\section*{ Acknowledgements}
\noindent
U. Rozikov thanks Aix-Marseille University, Institute for Advanced Study IM\'eRA (Marseille, France) for support by a residency
scheme. We thank the referee for a careful reading of the manuscript and helpful suggestions which have improved the paper.

\end{document}